# Fundamental gaps of finite systems from the eigenvalues of a generalized Kohn-Sham method

Tamar Stein[1], Helen Eisenberg[1], Leeor Kronik[2*], and Roi Baer[1*]

[1]*Fritz Haber Center for Molecular Dynamics, Institute of Chemistry, Hebrew University, Jerusalem 91904 Israel.*
[2]*Department of Materials and Interfaces, Weizmann Institute of Science, Rehovoth 76100, Israel*

We present a broadly-applicable, physically-motivated first-principles approach to determining the fundamental gap of finite systems. The approach is based on using a range-separated hybrid functional within the generalized Kohn-Sham approach to density functional theory. Its key element is the choice of a range-separation parameter such that Koopmans' theorem for both the neutral and anionic is obeyed as closely as possible. We demonstrate the validity, accuracy, and advantages of this approach on first, second, and third row atoms, the oligoacene family of molecules, and a set of hydrogen-passivated silicon nanocrystals. This extends the quantitative usage of density functional theory to an area long believed to be outside its reach.

The Kohn-Sham (KS)[1] formulation of density functional theory (DFT)[2] has become the method of choice for ground state electronic-structure calculations across an unusually wide variety of fields in physics, chemistry, and materials science.[3] In this approach, the interacting-electron system is mapped into an equivalent non-interacting electron system possessing the same density, $n$, that is subject to a common local external potential in the form:[4]

$$\left(-\frac{\nabla^2}{2} + v_{\text{ext}}(r) + v_{\text{H}}([n];r) + v_{\text{xc}}([n];r)\right)\varphi_i(r) = \varepsilon_i \varphi_i(r), \quad (1)$$

where $v_{\text{ext}}(r)$ is the ion-electron potential, $v_{\text{H}}([n];r)$ is the Hartree potential, $v_{\text{xc}}([n];r)$ is the exchange-correlation potential, and $\varepsilon_i$ and $\varphi_i(r)$ are KS eigenvalues and orbitals, respectively. This mapping is exact in principle, but in practice only approximate forms for $v_{\text{xc}}([n];r)$ are available.

One well-known deficiency of the KS approach is that the KS gap, $E_{KS}$, defined as the difference between the lowest unoccupied and highest occupied KS eigenvalues, generally differs from the fundamental gap, $E_g$, defined as the difference between the ionization potential, $I$, and the electron affinity, $A$.[5] This is because $E_g = E_{KS} + \Delta_{xc}$,[6] where $\Delta_{xc}$ is the derivative discontinuity, i.e., the finite "jump" that the exchange-correlation potential exhibits as the particle number crosses the integer number of particles in the system, $N$.[7, 8] While in some cases $\Delta_{xc}$ can be small, extensive numerical investigations show that it is usually sizable.[9] This would hinder the prediction of the fundamental gap from the KS eigenvalues of the $N$ electron system even if the exact form for $v_{\text{xc}}([n];r)$ were known.

A possible remedy for this deficiency may be found within the generalized Kohn-Sham (GKS) scheme,[5, 10, 11] in which the interacting-electron system is mapped into an interacting model system that can still be represented by a single Slater determinant. This leads to the GKS equation,

$$\left(\hat{O}_s[\{\varphi_i\}] + v_{\text{ext}}(r) + v_{\text{R}}([n];r)\right)\varphi_j(r) = \varepsilon_j \varphi_j(r), \quad (2)$$

where $\hat{O}_s[\{\varphi_i\}]$ is a generally non-local, orbital-specific operator and $v_{\text{R}}([n];r)$ is a "remainder" local potential, which includes all Hartree, exchange, correlation, or kinetic energy components not accounted for by $\hat{O}_s[\{\varphi_i\}]$. As in the KS theory, this mapping is exact in principle but approximate in practice. It is generally hoped that because $\hat{O}_s[\{\varphi_i\}]$ inherently exhibits a discontinuity as the particle number crosses an integer due to the partial occupation of an additional orbital, a judicious choice of $\hat{O}_s[\{\varphi_i\}]$ would greatly diminish the discontinuity in $v_R(r)$, making the GKS gap, $E_{GKS}$, considerably closer to $E_g$ than $E_{KS}$.[5, 10, 12] It is further hoped that the non-local character of $\hat{O}_s[\{\varphi_i\}]$ would allow it to mimic more efficiently the role played by the self-energy operator in many-body perturbation theory calculations of quasi-particle excitation energies, again suggesting that $E_{GKS}$ could be more suitable then $E_{KS}$ for predicting $E_g$.[13]

For solids, this hope is supported, to an extent, by practical computation schemes: In the screened-exchange approach,[10, 14] $\hat{O}_s[\{\varphi_i\}]$ corresponds to the sum of the single-particle kinetic energy operator, the Hartree operator, and a Fock-like operator based on a semi-classically screened potential. In the hybrid functional approach (which can also be viewed as a special case of the GKS scheme),[5] a weighted mixture of local exchange and non-local Fock exchange is used. In many (though not all) cases, either approach leads to meaningful improvement over the KS scheme in the prediction of fundamental gaps.[10, 14, 15]

For finite-sized objects (e.g., atoms, molecules, clusters, nanocrystals, etc.), these practical schemes are not useful for fundamental gap predictions. This is because the asymptotic potential, which is absent in a infinite solid, plays a crucial role in the energetics of electron addition and removal, i.e., in determining $I$ and $A$. In the screened exchange approach, the long-range exchange term is entirely absent. In conventional hybrid functionals, only a fixed fraction of it remains. And indeed (although the theoretical underpinnings and generality of this are still debated), it is often found that hybrid functional gaps for finite-sized objects are much smaller than the fundamental gap and are in far better agreement with the first excitation energy (also known as the optical gap).[5, 16]

Range-separated hybrid (RSH) functionals are a novel class of functionals, in which the exchange energy term is split into long-range and short-range terms, e.g., via $r^{-1} = r^{-1}\text{erf}(\gamma r) + r^{-1}\text{erfc}(\gamma r)$, or a similar use of the Yukawa potential.[17-19] The short-range exchange is represented by a local potential derived from the local-density or the generalized-gradient approximations, whereas the long-range part is treated via an "explicit" or "exact" exchange term. In this way, the full long-range Fock exchange can be obtained, without sacrificing the description of the short-range correlation that is essential to retaining a sufficiently accurate description of total energies.

If one assumes that an appropriate choice for γ is system independent, its value can be optimized for, e.g., thermochemistry and possibly other desired properties of the system or the functional.[18, 20-22] Specifically, Cohen et al. have shown that for such a semi-empirical RSH functional that they constructed, called MCY3,[22] GKS gaps for several atoms and small molecules were in good agreement with experimental fundamental gaps, with a mean absolute error of 0.7 eV.[12]

Assuming a system-independent γ is, however, only an approximation. A rigorous analysis, based on the adiabatic connection theorem, shows that in fact γ is itself a functional of the electron density, $n$.[19] For the homogeneous electron gas, Monte Carlo simulations show conclusively that $\gamma(n)$ strongly depends on the density.[20, 23] Furthermore, system-specific studies of γ showed that good prediction of, e.g., the ionization potential is possible, but that γ can vary substantially (with all else being equal) – from 0.3 for $Li_2$ to 0.7 for, e.g., HF or $O_2$ (γ values are given in atomic units throughout).[20] In particular, an important arena in which a constant γ is expected to be a problematic approximation is the study of the quantum size effect, i.e., the gap dependence on system size. This is because, in light of the above arguments, as the system evolves from the molecular limit to the solid-state limit, the relative importance of the long-range and short-range exchange must vary.

In this Letter, we show that with the aid of a simple, physically motivated, first principles γ-determining step, the GKS eigenvalues of RSH functionals can be used successfully for quantitative prediction of fundamental gaps of finite-sized objects in general and of quantum size effects in particular. This paves the road towards using DFT as a practical tool in an area dominated by computationally challenging methods such as coupled cluster, quantum Monte Carlo, or many-body perturbation theory calculations.

In exact KS theory, the DFT version of Koopmans' theorem establishes that the highest occupied KS eigenvalue is equal and opposite to the ionization potential.[7, 24] The starting point of our analysis is the realization that the same is true in exact GKS theory.[12] It implies that for obtaining the ionization potential from the highest-occupied GKS eigenvalue, an optimal choice for γ is to enforce Koopmans' theorem, i.e., to seek a value of γ such that

$$-\varepsilon_{HOMO}^{\gamma} = I^{\gamma}(N) \equiv E_{gs}(N-1;\gamma) - E_{gs}(N;\gamma), \quad (3)$$

where $\varepsilon_{HOMO}^{\gamma}$ is the highest occupied molecular orbital (HOMO) for a specific choice of γ and $I^{\gamma}(N)$ is the energy difference between the ground state energies, $E_{gs}$, of the $N$ and the $N-1$ electron system, per the same γ. While this has indeed been shown to be useful for determining ionization potentials in practice,[25] determining the gap requires that we also know the electron affinity. This means that we must employ Koopmans' theorem also for $I$ of the N+1 electron system which, barring relaxation effects, is the same as the $A$ of the $N$ electron system. Because there is only one parameter but two conditions, we now seek the γ that minimizes the overall deviation expressed in the target function[26, 27]

$$J(\gamma) = |\varepsilon_{HOMO}^{\gamma}(N) + I^{\gamma}(N)| + |\varepsilon_{HOMO}^{\gamma}(N+1) + I^{\gamma}(N+1)|. \quad (4)$$

Importantly, using Eq. (4) to choose the optimal γ, which we denote as γ*, does not require any empirical input and contains no adjustable parameters. Furthermore, two figures of merit can serve to evaluate if the result is expected to yield a usefully accurate fundamental gap. First, $J(\gamma^*)$ is expected to be substantially smaller than the desired accuracy. Second, one could demand a condition similar to Eq. (4) involving the lowest unoccupied molecular orbital (LUMO), in the form:[27]

$$J'(\gamma') = |\varepsilon_{HOMO}^{\gamma'}(N) + I^{\gamma'}(N)| + |\varepsilon_{LUMO}^{\gamma'}(N) + I^{\gamma'}(N+1)|. \quad (5)$$

Unlike Eq. (4), Eq. (5) has no rigorous basis because there is no formal equivalent to Koopmans' theorem involving the LUMO, owing to the derivative discontinuity. However, if indeed the residual derivative discontinuity is small, as hoped for above, then γ* should be close to γ'* and $J(\gamma^*)$ should be close to $J'(\gamma'^*)$.

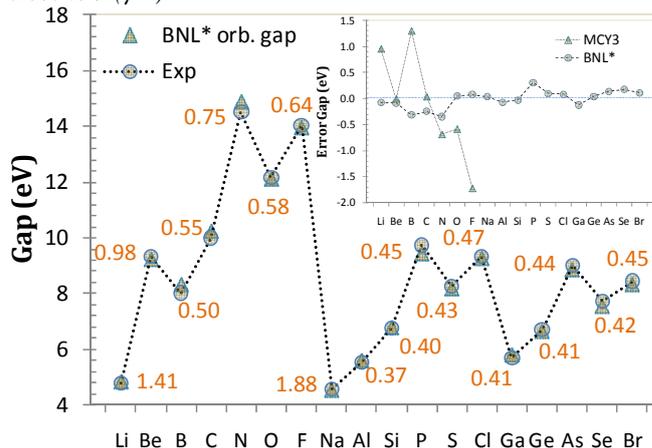

Figure 1: GKS/BNL* HOMO-LUMO gaps (computed using the aug-cc-pVTZ basis set), compared with experimental fundamental gaps.[28] The value of γ (in $a_0^{-1}$), determined by minimizing $J$, is indicated near each point. Inset: the deviation from experiment of GKS HOMO-LUMO gaps based on BNL* (this work) and MCY3[12].

To examine how well the generalized Kohn-Sham eigenvalues, obtained from such an optimally tuned RSH functional, can predict fundamental gaps in practice, we performed extensive optimally-tuned calculations based on the Baer, Neuhauser, and Livshits (BNL) range-separated hybrid functional, [20][29] as implemented in version 3.2 of Q-CHEM. [30]

We first consider a series of atoms from the first three rows of the periodic table. A comparison of computed and experimental gaps for these atoms is given in Figure 1. Several observations can be drawn from these results. First, the tuning procedure, based on Eq. (4) above, clearly produces consistently excellent gap prediction, with a mean deviation of -0.01 eV, a mean absolute deviation of 0.1 eV, and a maximal absolute deviation of 0.3eV (for phosphorus). Second, different atoms require different range-parameters (see Figure 1): For alkali metals γ is relatively high (between 1 and 2); this is because with a single valence electron, these atoms bear more resemblance to one-electron systems where Hartree-Fock theory is exact. With the exception of Be, the general trend along a given row in the periodic table is that higher values of the gap require higher values of γ. Moreover, excluding the alkali metals first row atoms require values of γ that vary between 0.5 and 0.65, whereas second and third row atoms require lower γ values, between 0.37 and 0.47. Third, for a given atom the gap is a very sensitive function of the range parameter. For example, for F and O the gap changes by as much as 8 eV when γ changes from 0.3 to 1. Fourth, the inset to Figure 1 compares the gaps obtained from our optimally tuned BNL functional to those obtained from the semi-empirical, fixed-γ MCY3[12] functional. Even though the latter presents a huge improvement over previous attempts, the optimal tuning clearly leads to higher accuracy and consistency. Taken together, these four observations underline the importance and non-triviality of our first-principles motivated tuning procedure.

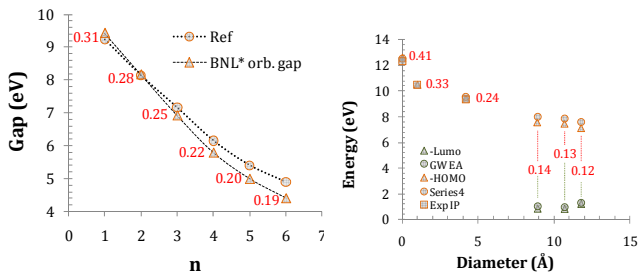

Figure 2: Left: GKS/BNL* HOMO-LUMO gaps, compared with gaps computed from experimental (vertical) ionization potentials[31] and best estimates of vertical electron affinities[32], for the polycyclic aromatic hydrocarbon linear oligomers, $C_{2+4n}H_{4+2n}$, with $n=1$ (benzene) to 6 (hexacene). The value of γ, determined by minimizing $J$, is indicated near each point. Right: GKS/BNL* HOMO and LUMO energies compared to GW[33] and experimental[34] ionization potentials (IP) and electron affinities (EA) of hydrogen terminated nano-crystalline spherical silicon fragments, as function of diameter. The values of the tuned range parameter are shown in red. In both systems the cc-pVTZ basis set was used. Geometries were obtained from a B3LYP calculation for the oligoacenes and from ref.[35] for the Si nanocrystals.

To test the power of our approach for molecules and nano-crystals, we studied the evolution of the fundamental gap in two families of systems that are well-known to exhibit a significant quantum size effect. One family is the polycyclic aromatic hydrocarbon linear oligomers - the oligoacene molecules $C_{2+4n}H_{4+2n}$ (n=1-6).[31, 32, 36, 37] Gap values obtained from total energy differences using B3LYP were found to be smaller than experiment by an average of 0.5 eV, with a maximal deviation for hexacene that exceeds $0.7\,eV$.[37] The orbital gaps of GKS/BNL*, compared with suitable reference values, are given in Figure 2 (left). The mean deviation from the reference of the BNL* orbital gap is $0.2\,eV$ and the mean absolute deviation is 0.3eV, with a maximal deviation of $0.5\,eV$. The tuned γ values decrease consistently with system size, from 0.3 $a_0^{-1}$ to 0.19 $a_0^{-1}$. This is physically reasonable: electron delocalization increases with system size, rendering the necessary weight of exact exchange smaller. Thus, it is again abundantly clear that an optimally-tuned γ will outdo a fixed one.

The other family we studied is the hydrogen-passivated spherical Si nanocrystals.[33, 38] In Figure 2 (right) the GKS/BNL* HOMO and LUMO energies for this system are compared with experimental[34] ionization potentials and GW-computed ionization potentials and electron affinities.[33]. The first three small-diameter systems do not bind an electron and the electron affinity is taken as zero when computing the gaps. The mean deviation of the BNL* eigenvalue gaps from GW gaps is 0.1 eV and the mean absolute deviation is 0.2eV with a maximal deviation of 0.45 eV. Moreover, $I$ and $A$ values are very well-reproduced separately, at a fraction of the computational cost of a GW calculation. As for the oligoacenes, here too the value of the range parameter decreases steadily as the nanocrystal size increases, and for the same physical reason.

We note that for the molecular and nanocrystaline systems, the remaining difference between our results and the reference values may also reflect limitations of the reference. For molecules, vertical electron affinities are hard to come by owing to structural relaxation effects. Furthermore, GW calculations of molecules may exhibit some deviation from experiment, especially for the electron affinity.[39] For example, the experimental ionization potential of $Si_5H_{12}$ is closer to the BNL* value than to the GW one (deviations of -0.01 and +0.15 eV, respectively).

Having demonstrated the accuracy and power of our approach, we revisit the above-discussed two figures of merit for its performance. For both oligoacenes and Si nanocrystals, the tuning procedure results in $J(\gamma^*)$ and $J'(\gamma'^*)$ that are close to zero (~$0.02\,eV$ on average), also indicating a negligible derivative discontinuity for these systems. Likely this excellent performance arises from the fact that addition of an electron to the system does not change its chemical nature, and ergo the required $\gamma^*$ for neutral and anion is similar. For atoms, addition of a single electron does change their nature appreciably. Coupled with the larger gaps in general, we expect to find larger deviations in this case. Indeed, optimal tuning based on Eq. (4) leads to an average root mean square deviation of $A$ from experiment of 0.15 eV. Furthermore, tuning based on Eq. (4) or (5) yields an average $J(\gamma^*)$ or $J'(\gamma^{*\prime})$ of 0.65 or 0.4 eV, respectively. The difference indicates a non-negligible, but still small, derivative discontinuity. Importantly, $J(\gamma^*)$ in this case is too strict a criterion. The remaining error in $I$ and $A$ is of the same sign, resulting in the excellent gaps of Fig. 1. Thus, even in this worst-case scenario, the method yields quantitatively useful fundamental gaps.

Finally, we comment on the importance of our approach. For finite systems, in principle one can always compute the fundamental gap from the total energy of the $N$, $N-1$, and $N+1$ electron systems. Depending on the system and the approximate functional, such calculations may be of insufficient accuracy in practice, e.g., for semi-local functionals applied to large systems.[40] But an accurate *eigenvalue-based* DFT gap is also essential in other contexts. Three typical examples are: (1) The incorrect DFT gap results in gross overestimation of the calculated conductance, e.g. of single-molecule benzenediamine−gold junctions.[41] (2) A more accurate DFT gap provides a much better starting point for DFT-based many-body perturbation theory calculations that yield the overall quasi-particle spectrum.[42] (3) Accurate DFT-level prediction of the fundamental gap is essential for accurate prediction of optically-induced charge transfer excitations.[26, 27]

In conclusion, we have presented a broadly-applicable, physically-motivated first-principles approach to determining the fundamental gap of finite systems. It is based on using a range-separated hybrid functional within the generalized Kohn-Sham scheme, while choosing the range-separation parameter such that Koopmans' theorem for both neutral and anion is obeyed as closely as possible. We demonstrated the validity, accuracy, and advantages of this approach on first, second, and third row atoms, the oligoacene family of molecules, and a set of hydrogen-passivated silicon nanocrystals. This extends the quantitative usage of density functional theory to an area long believed to be outside its reach.

This work was supported by the Israel Science Foundation.